\documentclass[journal ]{new-aiaa}
\usepackage[utf8]{inputenc}
\usepackage{textcomp}

\usepackage{acronym}
\usepackage[linesnumbered,ruled]{algorithm2e}
\usepackage{booktabs}
\usepackage{graphicx}
\usepackage{amsmath}
\usepackage[version=4]{mhchem}
\usepackage{siunitx}
\usepackage{standalone}
\usepackage{tikz} %
\usepackage{longtable,tabularx}
\acrodef{cpd}[CPD]{canonical polyadic decomposition}
\acrodef{tse}[TSE]{Taylor series expansion}
\title{Covariance Square Root Second-Order Mapping}

\author{Keith A.\ LeGrand \footnote{Assistant Professor, School of Aeronautics and Astronautics, klegrand@purdue.edu}}
\affil{Purdue University, West Lafayette, Indiana, 47907}
\author{Braden Hastings \footnote{PhD Student, Mechanical and Aerospace Engineering}, Jackson Kulik \footnote{Assistant Professor, Mechanical and Aerospace Engineering}}
\affil{Utah State University, Logan, Utah, 84322}

\newif\ifshownotes
\shownotestrue %
\newif\ifshowrevisionnotes
\showrevisionnotesfalse %

\begin{document}

\maketitle

\section{Introduction}
In Kalman filtering algorithms, numerical errors due to finite computer wordlength can lead to inconsistent state estimates, filter divergence, and total algorithmic failure.
In particular, while the covariance matrix is theoretically guaranteed to be symmetric positive semi-definite, numerical roundoff errors can lead to asymmetry and negative eigenvalues.
Such errors are unacceptable in safety-critical tracking and navigation applications, motivating alternative algorithms.

Inconsistencies between the theoretical and computational were identified as early as the Apollo program, when James Potter showed that numerical issues could be mitigated by square root factorizations of the covariance matrix.
Potter subsequently introduced the first square root filter, which was restricted to systems with no process noise and scalar measurement updates \cite{battin1964AstronauticalGuidance,maybeck1979StochasticModelsEstimation}.
Soon after, Schmidt identified a method for incorporating process noise in the square root propagation using triangularization \cite{schmidt1970ComputationalTechniquesKalman}.
Carlson introduced a method that leverages upper-triangular Cholesky factors of the covariance matrix, improving both speed and memory utilization over Potter's method.
Like Potter's method, Carlson's method processes vector measurements iteratively as scalars, albeit using rank-one modifications of the upper Cholesky factor \cite{carlson1973FastTriangularFormulation}.
The Carlson measurement update and Schmidt propagation together compose the Carlson-Schmidt square root filter, which operates entirely on Cholesky factors and thus avoids intermediate full covariance computations and the associated loss of precision.
The two methods underpinning Carlson-Schmidt---namely, matrix triangularization and rank-one modifications---have more recently been leveraged to establish square root formulations of the unscented Kalman filter \cite{vandermerwe2001SquarerootUnscentedKalman}.
Other work has proposed hyperbolic householder reflections to establish square root variants of the consider Kalman filter \cite{mccabe2018SquareRootConsiderFilters}.
Notably, existing square root methods focus on linear covariance and cubature-based mapping.

In highly nonlinear systems, higher-order covariance mapping has shown to produce significantly more accurate results than linear covariance mapping \cite{park2006nonlinear,siciliano2026HigherOrderTensorBasedDeferral}.
In many cases, higher-order Taylor coefficients of the dynamics solution flow and measurement map may be available in closed form or be computed already for other reasons and thus opportunistically available for negligible or no additional computational cost.
This article presents a Cholesky square root formulation of second-order covariance mapping for numerical stability and accuracy.
The resulting method finds applications in uncertainty quantification, navigation, tracking, and other state estimation applications in nonlinear systems.
A key challenge of factoring the fourth central moment tensor is solved by leveraging a \ac{cpd} of the fourth-order identity tensor as was first performed in order to speed up second-order covariance mapping \cite{hastings25}.
The numerical accuracy of the proposed method is compared to existing full covariance-based computations in two distinct nonlinear transformations.
The new square root method demonstrates favorable performance compared to its full covariance counterpart, producing orders of magnitude lower errors.
The remainder of this article is organized as follows.
Background information on square root factorization and triangularization is reviewed in Section~\ref{sec:background}.
The square root second-order covariance mapping method is developed in Section~\ref{sec:methodology}.
Numerical experiments are discussed in Section~\ref{sec:numerical_experiments}, and conclusions are summarized in Section~\ref{sec:conclusion}.

\section{Background}%
\label{sec:background}
\subsection{Matrix Square Roots and the Cholesky Decomposition}
The square root of a matrix has two common definitions in the literature.
In some literature, a matrix square root $\mathbf{A}^{1/2}$ of a given matrix $\mathbf{A}$ is any matrix such that $\mathbf{A}=\mathbf{A}^{1/2}\mathbf{A}^{1/2}$. However, in the context of state estimation, Kalman filtering, and statistics more broadly \cite{kessy2018optimal}, the matrix square root is often defined alternatively as
\begin{equation}
  \label{eq:matrix_sqrt_definition}
    \mathbf{A}=\mathbf{A}^{1/2}(\mathbf{A}^{1/2})^\top
\end{equation}
Equation~(\ref{eq:matrix_sqrt_definition}) is the definition of the matrix square root adopted in this paper.
The matrix square root is not unique since right multiplication of any matrix square root by a matrix with orthonormal rows results in another matrix square root.
However, one particularly useful matrix square root is the Cholesky factor
\begin{equation}
    \mathbf{A}=\mathbf{S}\mathbf{S}^\top
\end{equation}
where the Cholesky factor $\mathbf{S}$ is the unique lower triangular matrix square root with positive entries along the diagonal.
This form is useful because the Cholesky factorization is efficient, stable, and widely implemented.
In practice, its triangularity also allows storage cost reduction since only the lower half of the matrix is nonzero.
Further, this particular matrix square root admits efficient and stable rank-1 update/downdate algorithms.
Throughout this work, the algorithm developed by Stewart \cite{stewart1979EffectsRoundingError} is adopted for efficient rank-1 downdates, modified to operate on lower triangular Cholesky factors.
Given a lower Cholesky factor $\mathbf{S}$ and vector $\mathbf{v}$, the downdate operation is abbreviated as
\begin{align}
  \mathbf{B} = \texttt{choldowndate}(\mathbf{S}, \mathbf{v})
\end{align}
where $\mathbf{B}$ is the lower triangular Cholesky factor of the matrix
\begin{align}
  \mathbf{B} \mathbf{B}^{\top} = \mathbf{S}\mathbf{S}^{\top} - \mathbf{v} \mathbf{v}^{\top}
\end{align}

\subsection{The QR Factorization}
The QR factorization--computed via Householder reflections, Givens rotations, or by the modified Gram-Schmidt process \cite{golub2013matrix}--can also be useful as a tool to bring a generic matrix square root into lower triangular form.
In particular, suppose that an arbitrary, potentially rectangular matrix square root~$\mathbf{A}^{1/2}$ is known.
The Cholesky factor of $\mathbf{A}$ can be computed from
\begin{align}
    &(\mathbf{A}^{1/2})^\top=\mathbf{Q}\mathbf{R}\implies \mathbf{A}=\mathbf{R}^\top\mathbf{Q}^\top\mathbf{Q}\mathbf{R}=\mathbf{R}^\top\mathbf{R}
\end{align}
where $\mathbf{Q}$ is a matrix with orthogonal columns and~$\mathbf{R}$ is an upper triangular matrix with positive diagonal entries by most conventions.
Clearly, given the QR factorization of the transpose of the matrix square root, the transpose of the resulting~$\mathbf{R}$ factor is the Cholesky factor of the original matrix $\mathbf{A}$.
For the rest of this paper, the shorthand~$\texttt{qr}_\mathbf{R}(\mathbf{B})$ will be used to denote the triangular factor~$\mathbf{R}$ from the QR decomposition of the matrix $\mathbf{B}$.

\section{Methodology}
\label{sec:methodology}
Given a random variable $\mathbf{x}\sim \mathcal{N}(\mathbf{m}_x, \mathbf{P}_x)$, noise described by the matrix $\boldsymbol{\Gamma}\in\mathbb{R}^{m\times n}$ as well as the random vector $\mathbf{w}\sim \mathcal{N}(\mathbf{0}, \mathbf{P}_w)$, and a nonlinear function $\mathbf{g}: \mathbb{R}^{n}\rightarrow\mathbb{R}^m$, an approximation of $\mathbf{z}=\mathbf{g}(\mathbf{x}) + \boldsymbol{\Gamma}\mathbf{w}$ via its first two moments is sought---often so that $\mathbf{z}$ may be approximated as a Gaussian random variable with those two moments.
In particular, this paper employs a second-order approximation of the nonlinear function in deriving an approximation of the first two moments:
\begin{equation}
    g^i(\mathbf{m}_x+\delta\mathbf{x})\approx g^i(\mathbf{m}_x)+\frac{\partial g^i}{\partial x^j}\bigg\vert_{\mathbf{m}_x}\delta x^j+\frac{1}{2}\frac{\partial g^i}{\partial x^j \partial x^k}\bigg\vert_{\mathbf{m}_x}\delta x^j\delta x^k
\end{equation}
where the Einstein summation convention is employed here and throughout the remainder of the paper.

\subsection{First-Order Covariance Square Root}%
\label{sec:first_order_covariance_square_root}
When only the terms up to first-order are retained, the familiar mean and covariance mapping equations are obtained:
\begin{align}
  \label{eq:first_order_mean}
   \mathrm{E}[\mathbf{z}] &\approx \mathbf{m}_{(1)} = \mathbf{g}(\mathbf{m}_{x})\\
   \label{eq:first_order_cov_prop_full}
   \mathrm{E}[(\mathbf{z} - \mathbf{m}_{(1)})(\mathbf{z} - \mathbf{m}_{(1)})^{\top}] &\approx  \mathbf{P}_{(1)} = \mathbf{G}^{(1)} \mathbf{P}_{x} (\mathbf{G}^{(1)})^{\top} + \boldsymbol{\Gamma} \mathbf{P}_{w}  \boldsymbol{\Gamma}^{\top}
\end{align}
where the Jacobian of the nonlinear function evaluated at the mean is defined as~$\mathbf{G}^{(1)} = \frac{\partial \mathbf{g}}{\partial \mathbf{x}}\big|_{\mathbf{x}=\mathbf{m}_{x}}$.
Denote by~$\mathbf{S}_{x}$ and~$\mathbf{S}_{w}$ the lower triangular Cholesky factors of the covariance matrices $\mathbf{P}_{x}$ and $\mathbf{P}_{w}$ respectively such that $\mathbf{P}_{[\cdot]} = \mathbf{S}_{[\cdot]} \mathbf{S}_{[\cdot]}^{\top}$. %
By substituting the known input square root factors in~\eqref{eq:first_order_cov_prop_full} and noting its symmetric form, a block matrix factorization emerges as
\begin{align}
  \label{eq:first_order_square_root_rectangular}
 \mathbf{S}_{(1)} \mathbf{S}_{(1)}^{\top}
 =
 [\mathbf{G}^{(1)} \mathbf{S}_{x} \,\, \mid \,\, \boldsymbol{\Gamma}\mathbf{S}_{w}]
 [\mathbf{G}^{(1)} \mathbf{S}_{x} \,\, \mid \,\, \boldsymbol{\Gamma}\mathbf{S}_{w}]^{\top}
\end{align}
where the block-partitioned matrix on the right hand side is a valid, albeit rectangular, square root factor of the propagated covariance $\mathbf{P}_{(1)}$.
Thus, the lower Cholesky factor of $\mathbf{P}_{(1)}=\mathbf{S}_{(1)}\mathbf{S}_{(1)}^{\top}$ is readily obtained by means of QR decomposition as
\begin{align}
  \label{eq:first_order_cov_prop_sqrt}
  \mathbf{S}_{(1)} = \texttt{qr}_\mathbf{R}([\mathbf{G}^{(1)}\mathbf{S}_{x} \,\, \mid \,\, \boldsymbol{\Gamma} \mathbf{S}_{w}]^{\top})^{\top}
\end{align}
\subsection{Second-Order Covariance Square Root}%
\label{sec:second_order_covariance_square_root}

A higher accuracy approximation of the mean and covariance can be obtained by retaining higher-order terms in the \ac{tse} of $\mathbf{g}$ about $\mathbf{m}_{x}$.
Given a second-order \ac{tse} approximation of $\mathbf{g}$, then the associated second-order approximation of the propagated mean and covariance is
\begin{align}
  \mathrm{E}[\mathbf{z}]
  &\approx
  \mathbf{m}_{(2)}
  =
  \mathbf{g}(\mathbf{m}_x)+\boldsymbol{\delta} \mathbf{m}_{(2)}\\
  \label{eq:second_order_covariance}
  \mathrm{E}[(\mathbf{z}-\mathrm{E}[\mathbf{z}])(\mathbf{z}-\mathrm{E}[\mathbf{z}])]
  &\approx
  \mathbf{P}_{(2)}
  =
  \mathbf{P}_{(1)} + \boldsymbol{\delta} \mathbf{P}_{(2)} - \boldsymbol{\delta}\mathbf{m}_{(2)}\delta\mathbf{m}_{(2)}^{\top} + \mathbf{P}_{w}\\
  (\boldsymbol{\delta} \mathbf{m}_{(2)})^i
  &=
  \frac{1}{2} (\mathbf{G}^{(2)})^{i}_{jk} (\mathbf{P}_x)^{jk}\\
  \label{eq:second_order_covariance_correction}
  (\boldsymbol{\delta} \mathbf{P}_{(2)})^{ij}
  &=
  \frac{1}{4}
  (\mathbf{G}^{(2)})^{i}_{kl}
  (\mathbf{G}^{(2)})^{j}_{pq}
  \mathrm{E}[(\mathbf{x}-\mathbf{m}_{x})^k(\mathbf{x}-\mathbf{m}_{x})^l(\mathbf{x}-\mathbf{m}_{x})^p(\mathbf{x}-\mathbf{m}_{x})^q]
\end{align}
which makes no assumption of $\mathbf{x}$ being Gaussian \cite{park2006nonlinear}.
The vector $\boldsymbol{\delta}{\mathbf{m}}_{(2)}$ can be viewed as the second-order correction factor to the first-order approximation of the mean, $\mathbf{m}_{(1)}$.
In the same way, the matrix $\boldsymbol{\delta}\mathbf{P}_{(2)}$ can be interpreted as a second-order correction to the first-order covariance approximation and is a function of the fourth central moment of $\mathbf{x}$, as seen in~\eqref{eq:second_order_covariance_correction}.
In the special case that $\mathbf{x}$ is Gaussian distributed, its fourth central moment can be expressed explicitly in terms of its covariance by Isserlis's theorem as
\begin{align}
  \mathrm{E}[(\mathbf{x}-\mathbf{m}_{x})^k(\mathbf{x}-\mathbf{m}_{x})^l(\mathbf{x}-\mathbf{m}_{x})^p(\mathbf{x}-\mathbf{m}_{x})^q]
  &=
  (\mathbf{P}_{x})^{kl}
  (\mathbf{P}_{x})^{pq}
  +
  (\mathbf{P}_{x})^{kp}
  (\mathbf{P}_{x})^{lq}
  +
  (\mathbf{P}_{x})^{kq}
  (\mathbf{P}_{x})^{lp}
\end{align}
Thus, in this case, the second-order term $\boldsymbol{\delta}\mathbf{P}_{(2)}$ can be expressed in terms of the input covariance as
\begin{align}
  \label{eq:second_order_covariance_correction_gaussian}
  (\boldsymbol{\delta} \mathbf{P}_{(2)})^{ij}
  &=
  \frac{1}{4}
  (\mathbf{G}^{(2)})^{i}_{kl}
  (\mathbf{G}^{(2)})^{j}_{pq}
  \left[
  (\mathbf{P}_{x})^{kl}
  (\mathbf{P}_{x})^{pq}
  +
  (\mathbf{P}_{x})^{kp}
  (\mathbf{P}_{x})^{lq}
  +
  (\mathbf{P}_{x})^{kq}
  (\mathbf{P}_{x})^{lp}
  \right]
\end{align}

Our objective is to obtain a lower square root factor of $\mathbf{P}_{(2)}$ such that $\mathbf{P}_{(2)} = \mathbf{S}_{(2)}\mathbf{S}_{(2)}^{\top}$.
Assume for the moment that the square root $\boldsymbol{\delta} \mathbf{S}_{(2)}$ of $\boldsymbol{\delta}\mathbf{P}_{(2)}$ is available.
Then
\begin{align}
  \label{eq:second_order_covariance_s1s2_factored}
  \mathbf{S}_{(2)}
  \mathbf{S}_{(2)}^{\top}
  =
  \big[
    \mathbf{G}_{(1)} \mathbf{S}_{x} \,\,\mid\,\, \boldsymbol{\delta}\mathbf{S}_{(2)} \,\,\mid\,\, \boldsymbol{\Gamma} \mathbf{S}_{w}
  \big]
  \big[
    \mathbf{G}_{(1)} \mathbf{S}_{x} \,\,\mid \,\, \boldsymbol{\delta}\mathbf{S}_{(2)} \,\,\mid\,\, \boldsymbol{\Gamma} \mathbf{S}_{w}
  \big]^{\top}
  -
  \boldsymbol{\delta}\mathbf{m}_{(2)}\delta\mathbf{m}_{(2)}^{\top}
\end{align}
From~\eqref{eq:second_order_covariance_s1s2_factored}, it is clear that, once the factor $\boldsymbol{\delta}\mathbf{S}_{(2)}$ is known, then $\mathbf{S}_{(2)}$ can be obtained by performing a QR factorization of the rectangular factor, followed by a Cholesky downdate by the second-order mean correction $\boldsymbol{\delta}\mathbf{m}_{(2)}$.

The challenge of second-order square root covariance mapping lies primarily in obtaining a square root factor~$\boldsymbol{\delta}\mathbf{S}_{(2)}$ without needing to first compute~$\boldsymbol{\delta}\mathbf{P}_{(2)}$.
In this case, no simple rectangular factorization similar to~\eqref{eq:first_order_square_root_rectangular} is available.
This is because the fourth-order central moment tensor couples the two second-order partial derivative terms in~\eqref{eq:second_order_covariance_correction} in a nontrivial way.
The rest of this section focuses on the factorization of~$\boldsymbol{\delta}\mathbf{P}_{(2)}$, which leverages the identity tensor and its \ac{cpd}.

Consider first a standard multivariate Gaussian random variable~$\mathbf{y}$, which is related to the estimation error~$\mathbf{x} - \mathrm{E}[\mathbf{x}]$ through a whitening transformation $\mathbf{y} = \mathbf{S}_{x}^{-1} (\mathbf{x} - \mathrm{E}[\mathbf{x}])$ such that its covariance is the identity matrix $\mathbf{I}^{(2)}$.
By Isserlis's theorem, the fourth central moment is a fourth-order tensor expressed in terms of the identity matrix as
\begin{align}
  \label{eq:standard_gaussian_fourth_moment}
  \mathrm{E}[y^{k} y^{l} y^{p} y^{q}]
  =
  (\mathbf{I}^{(2)})^{kl}
  (\mathbf{I}^{(2)})^{pq}
  +
  (\mathbf{I}^{(2)})^{kp}
  (\mathbf{I}^{(2)})^{lq}
  +
  (\mathbf{I}^{(2)})^{kq}
  (\mathbf{I}^{(2)})^{lp}
\end{align}
Equation~\eqref{eq:standard_gaussian_fourth_moment} exhibits multiple symmetries.
First, by symmetry of the identity matrix, $(\mathbf{I}^{(2)})^{kl} = (\mathbf{I}^{(2)})^{lk}$.
Second, the second two terms in~\eqref{eq:standard_gaussian_fourth_moment} are index permutations of the first term.
It then follows that
\begin{align}
  (\mathbf{I}^{(2)})^{kl}
  (\mathbf{I}^{(2)})^{pq}
  +
  (\mathbf{I}^{(2)})^{kp}
  (\mathbf{I}^{(2)})^{lq}
  +
  (\mathbf{I}^{(2)})^{kq}
  (\mathbf{I}^{(2)})^{lp}
  =
  3 \operatorname{sym}(\mathbf{I}^{(2)} \bigcirc \mathbf{I}^{(2)})^{klpq}
\end{align}
where $\operatorname{sym}(\cdot)$ is the tensor symmetrization operation, which averages a tensor over all index permutations, and $\bigcirc$ denotes the tensor outer product.
The symmetrization of the tensor product of identity matrices is sometimes known as the identity tensor.
It appears in the definition of the generalization of the characteristic polynomial for studying eigenvalues of tensors \cite{qi2005eigenvalues}.
Additionally, it has two variational characterizations.
It is the unique super-symmetric tensor that acts on four copies of any unit vector to produce unity
\begin{equation}
    \left(\mathbf{I}^{(4)}\right)_{ijkl}u^iu^ju^ku^l=1 \quad  \forall \quad \Vert\mathbf{u}\Vert_2=1
\end{equation}
and the unique super-symmetric tensor that acts on three copies of any unit vector to produce the same unit vector as output \cite{kolda2011shifted}
\begin{equation}
    \left(\mathbf{I}^{(4)}\right)_{ijkl}u^iu^ju^k=u_l \quad  \forall \quad \Vert\mathbf{u}\Vert_2=1
\end{equation}
It is simple to see that the symmetrization of the outer product of identity matrices satisfies both of these variational characterizations, and uniqueness comes from the fact that the only perturbative super-symmetric tensor that acts via four times repeated contraction on all unit vectors to produce zero is the tensor consisting of all zeros.
The construction of identity tensors of higher even order and their variational characterizations are essentially identical and are based around using a tensor product with additional identity matrices.
Thus, the fourth central moment of the standard Gaussian is three times the fourth-order identity tensor:
\begin{align}
  \mathrm{E}[\mathbf{y}^{\otimes 4}]^{klpq}
  =
  \mathrm{E}[y^{k} y^{l} y^{p} y^{q}]
  =
  3(\mathbf{I}^{(4)})^{klpq}
\end{align}

The fourth central moment of the arbitrary multivariate Gaussian $\mathbf{x}$ can now be related to the standard multivariate Gaussian.
By the inverse whitening transformation $\mathbf{x}  - \mathrm{E}[\mathbf{x}] = \mathbf{S}_{x} \mathbf{y}$, the fourth central moment of $\mathbf{x}$ is
\begin{align}
  \mathrm{E}[(\mathbf{x}-\mathbf{m}_{x})^k(\mathbf{x}-\mathbf{m}_{x})^l(\mathbf{x}-\mathbf{m}_{x})^p(\mathbf{x}-\mathbf{m}_{x})^q]
  &=
  (\mathbf{S}_{x})^{k}_{k'}
  (\mathbf{S}_{x})^{l}_{l'}
  (\mathbf{S}_{x})^{p}_{p'}
  (\mathbf{S}_{x})^{q}_{q'}
  \mathrm{E}[y^{k'}y^{l'}y^{p'}y^{q'}]\\
  \label{eq:fourth_moment_SSSS_3I}
  &=
  (\mathbf{S}_{x})^{k}_{k'}
  (\mathbf{S}_{x})^{l}_{l'}
  (\mathbf{S}_{x})^{p}_{p'}
  (\mathbf{S}_{x})^{q}_{q'}
  (3 \mathbf{I}^{(4)})^{k'l'p'q'}
\end{align}
Substitution of~\eqref{eq:fourth_moment_SSSS_3I} in~\eqref{eq:second_order_covariance_correction} yields an expression for~$\boldsymbol{\delta} \mathbf{P}_{(2)}$ that is expressed explicitly in terms of the square root factor~$\mathbf{S}_{x}$, the second-order partial derivative tensor~$\mathbf{G}^{(2)}$, and the fourth-order identity tensor~$\mathbf{I}^{(4)}$.
Yet, even in this form, no square root factorization of the final result emerges without further manipulation.

Factorization of~$\boldsymbol{\delta} \mathbf{P}_{(2)}$ is made possible by decomposing the fourth-order identity tensor.
The fourth-order identity tensor can be represented exactly by its symmetric \ac{cpd}
\begin{align}
   \mathbf{I}^{(4)}
   &=
   \sum_{r=1}^{R} w_{r} \boldsymbol{v}_{r}^{\otimes 4}\\
   (3\mathbf{I}^{(4)})^{klpq}
   &=
   \left(\sum_{r=1}^{R} \left[(3w_{r})^{\frac{1}{4}} \boldsymbol{v}_{r}\right]^{\otimes 4}\right)^{klpq}
   =
   \sum_{r=1}^{R}
     \beta_{r}
     \beta_{r}
     (\boldsymbol{v}_{r})^{k}
     (\boldsymbol{v}_{r})^{l}
     (\boldsymbol{v}_{r})^{p}
     (\boldsymbol{v}_{r})^{q}
\end{align}
where the factors $\mathbf{v}_{r}$ have unit norm, the scaled and square rooted weights $\beta_{r}=(3w_{r})^{\frac{1}{2}}$, and $R$ is the rank of $\mathbf{I}^{(4)}$, which depends on the dimension of $\mathbf{x}$.
The weights and factors depend only on the dimension of $\mathbf{x}$, and thus they can be computed once offline for a range of state dimensions and referenced at runtime.
For highest numerical accuracy and efficiency, the modified weights $\beta_{r}$ should be computed and stored at the highest possible numerical precision instead of the identity tensor \ac{cpd} weights, $w_{r}$, or their fourth roots.
The symmetric \ac{cpd} computation is detailed in Section~\ref{sec:symmetric_canonical_polyadic_decomposition}.

By employing this \ac{cpd} of the fourth-order identity tensor, the second-order covariance correction is expressed as
\begin{align}
  \label{eq:second_order_covariance_correction_cpd}
  (\boldsymbol{\delta} \mathbf{P}_{(2)})^{ij}
  &=
  \frac{1}{4}
  (\mathbf{G}^{(2)})^{i}_{kl}
  (\mathbf{G}^{(2)})^{j}_{pq}
  (\mathbf{S}_{x})^{k}_{k'}
  (\mathbf{S}_{x})^{l}_{l'}
  (\mathbf{S}_{x})^{p}_{p'}
  (\mathbf{S}_{x})^{q}_{q'}
  (3 \mathbf{I}^{(4)})^{k'l'p'q'}\\
  &=
  \sum_{r=1}^{R}
  \left(
    \frac{1}{2}
     \beta_{r}\cdot
    (\mathbf{G}^{(2)})^{i}_{kl}
    (\mathbf{S}_{x})^{k}_{k'}
    (\mathbf{S}_{x})^{l}_{l'}
     (\boldsymbol{v}_{r})^{k'}
     (\boldsymbol{v}_{r})^{l'}
  \right)
  \left(
    \frac{1}{2}
     \beta_{r}\cdot
    (\mathbf{G}^{(2)})^{j}_{pq}
    (\mathbf{S}_{x})^{p}_{p'}
    (\mathbf{S}_{x})^{q}_{q'}
     (\boldsymbol{v}_{r})^{p'}
     (\boldsymbol{v}_{r})^{q'}
  \right)\\
  &=
  \sum_{r=1}^{R}
  \mathbf{p}_{r} \mathbf{p}_{r}^{\top}
\end{align}
where
\begin{align}
  (\mathbf{p}_{r})^{i}
  =
    \frac{1}{2}
     \beta_{r}\cdot
    (\mathbf{G}^{(2)})^{i}_{kl}
    (\mathbf{S}_{x})^{k}_{k'}
    (\mathbf{S}_{x})^{l}_{l'}
     (\boldsymbol{v}_{r})^{k'}
     (\boldsymbol{v}_{r})^{l'}
\end{align}
Thus, it follows that a valid---yet rectangular---square root factor of $\boldsymbol{\delta}\mathbf{P}_{(2)}$ is
\begin{align}
  \boldsymbol{\delta}{\mathbf{S}}_{(2)} = [\mathbf{p}_{1} \,\, \cdots \,\, \mathbf{p}_{R}]
\end{align}
With this, the square root factor of the second-order propagated covariance matrix satisfies
\begin{align}
  \label{eq:second_order_covariance_s1s2_factored}
  \mathbf{S}_{(2)}
  \mathbf{S}_{(2)}^{\top}
  =
  \big[
    \mathbf{G}^{(1)} \mathbf{S}_{x} \,\,\mid\,\, \mathbf{p}_{1} \,\, \cdots \,\, \mathbf{p}_{R} \,\,\mid\,\, \boldsymbol{\Gamma} \mathbf{S}_{w}
  \big]
  \big[
    \mathbf{G}^{(1)} \mathbf{S}_{x} \,\,\mid\,\, \mathbf{p}_{1} \,\, \cdots \,\, \mathbf{p}_{R} \,\,\mid\,\, \boldsymbol{\Gamma} \mathbf{S}_{w}
  \big]^{\top}
  -
  \boldsymbol{\delta}\mathbf{m}_{(2)}\delta\mathbf{m}_{(2)}^{\top}
\end{align}
Numerically, the Cholesky factor of the second-order approximation of the covariance can be obtained by first taking the QR decomposition of the rectangular matrix on the right hand side of~\eqref{eq:second_order_covariance_s1s2_factored} and then performing a Cholesky rank-1 downdate based on the second-order perturbation to the mean:
\begin{align}
  \mathbf{S}_{(2)} &= \texttt{choldowndate}(\mathbf{A}, \boldsymbol{\delta}\mathbf{m}_{(2)})\\
   \mathbf{A} &= \texttt{qr}([\mathbf{G}^{(1)} \mathbf{S}_{x} \,\,\mid\,\, \mathbf{p}_{1} \,\, \cdots \,\, \mathbf{p}_{R} \,\,\mid\,\, \boldsymbol{\Gamma} \mathbf{S}_{w}]^{\top})^{\top}
\end{align}
Note that the Cholesky downdate is necessary, as the negative rank-1 perturbation of the matrix cannot be directly incorporated into the QR trick because of the sign.
Algorithm~\ref{alg:second_order_sqrt_prop} summarizes the overall second-order square root covariance mapping scheme, assuming that the \ac{cpd} rank $R$ factors $\mathbf{v}_r$ and modified weights $\beta_{r}$ have been precalculated for the fourth-order identity tensor associated with the dimension of the variable $\mathbf{x}$.

\begin{algorithm}[htbp]
    \caption{Second-Order Square Root Covariance Mapping}
    \label{alg:second_order_sqrt_prop}
    \KwIn{$\mathbf{S}_{x}, \, \mathbf{S}_{w}, \, \boldsymbol{\Gamma}, \, \mathbf{G}^{(1)}, \, \mathbf{G}^{(2)}$}
    \KwOut{$\mathbf{S}_{(2)}$}
    $(\boldsymbol{\delta} \mathbf{m}_{(2)})^i = \frac{1}{2} (\mathbf{G}^{(2)})^{i}_{jk} (\mathbf{S}_x)^{jl}(\mathbf{S}_x)^{kl}$\\
    \ForEach{$r=1,\ldots,R$}{
        $(\mathbf{u}_{r})^{i} = \left(\mathbf{S}_{x}\right)_{j}^i\left(\boldsymbol{v}_r\right)^{j}$\\
        $(\mathbf{p}_{r})^{i} = \frac{1}{2} \beta_{r} \cdot (\mathbf{G}^{(2)})_{k l}^i(\mathbf{u}_r)^k(\mathbf{u}_r)^l$
    }
    $\mathbf{A} \gets \texttt{qr}_{\mathbf{R}}([\mathbf{G}^{(1)}\mathbf{S}_{x} \,\,\mid\,\, \mathbf{p}_{1} \,\, \cdots \quad \mathbf{p}_{R} \,\,\mid\,\, \boldsymbol{\Gamma} \mathbf{S}_{w}]^{\top})^{\top}$\\
    $\mathbf{S}_{(2)} \gets \texttt{choldowndate}(\mathbf{A},\delta \mathbf{m}_{(2)})$
\end{algorithm}

Examining the standard approach to second-order covariance analysis without the square root factor, the computational complexity is dominated by the tensor contraction in~\eqref{eq:second_order_covariance_correction}, which is $\mathcal{O}(m^2n^4)$ in the number of basic floating point operations.
On the other hand, Algorithm~\ref{alg:second_order_sqrt_prop} has two operations that on initial examination may dominate the computational cost.
These are the formation of the factors $\mathbf{p}_r$ and then the computation of the $\mathbf{R}$ factor from the QR decomposition of a large rectangular matrix.
The computational complexity of forming all of the factors $\mathbf{p}_r$ is $\mathcal{O}(Rmn^2)$ while the computation of the $\mathbf{R}$ factor from the QR decomposition of a $(2n+R)\times m$ dimensional matrix is $\mathcal{O}(2m^2(2n+R-m/3))$ \cite{higham2008functions}.
Because $m<n$, the cost will typically be dominated by the calculation of the $\mathbf{p}_r$ factors.
Thus, a speedup from $\mathcal{O}(m^2n^4)$ to $\mathcal{O}(Rmn^2)$ is enjoyed assuming that $R$ is a small constant relative to $mn^2$, as is typically the case in those \acp{cpd} examined by the authors.

\subsection{Symmetric Canonical Polyadic Decomposition}%
\label{sec:symmetric_canonical_polyadic_decomposition}
The \ac{cpd} of a general tensor is often obtained via alternating least squares on each set of potentially different factor vectors (as opposed to the form of the \textit{symmetric} \ac{cpd} presented here which utilizes the same vector four times) \cite{kolda2009tensor}.
When seeking the symmetric \ac{cpd} of a super-symmetric tensor, existing techniques \cite{carroll1970analysis,faber2003recent} that neither assume symmetry of the tensor nor symmetry of the resulting decomposition can sometimes yield a symmetric \ac{cpd}, though this is not guaranteed.
A direct optimization approach has also been developed in order to guarantee symmetry of the final decomposition while also exploiting symmetry of the original tensor \cite{kolda2015numerical}.
This work employs the tensor toolbox alternating least squares formulation, which has resulted in symmetric \acp{cpd} in these cases of interest \cite{ttbox}.

\begin{figure}[htpb]
  \centering
  \begin{tikzpicture}
     \node[anchor=north west] (img) at (0,0)
    {\includegraphics[width=0.3\linewidth]{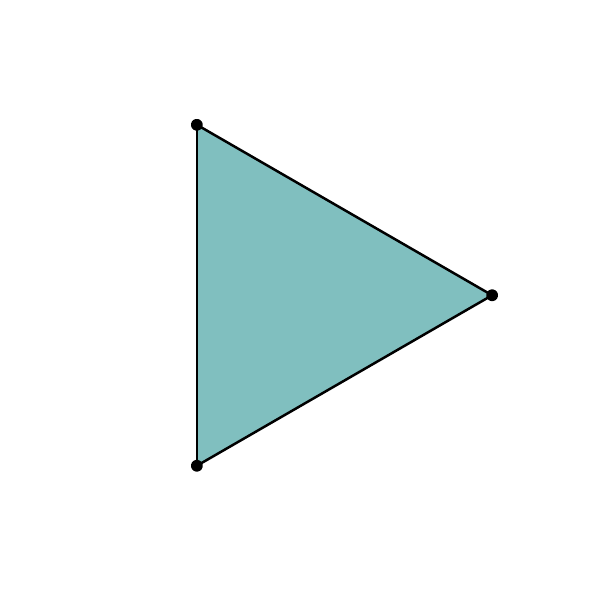}};
    \node [text=black]at (0.37,-0.85) {(a)};
  \end{tikzpicture}
  \begin{tikzpicture}
     \node[anchor=north west] (img) at (0,0)
    {\includegraphics[width=0.3\linewidth]{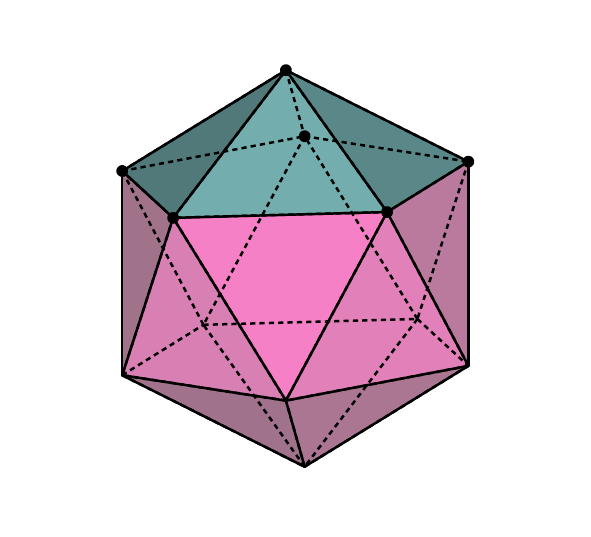}};
    \node [text=black]at (0.37,-0.3) {(b)};
  \end{tikzpicture}
  \caption{Symmetric canonical polyadic decomposition factors of the fourth-order identity tensor in dimensions two (a) and three (b).}%
  \label{fig:geometry}
\end{figure}

In some low-dimensional cases, the symmetric \ac{cpd} can be obtained analytically upon guessing the correct rank.
It seems that for the two dimensional identity tensor, the \ac{cpd} factors of the fourth-order identity tensor are given by the vertices of an equilateral triangle centered at the origin (oriented in any fashion), where the uniform weight across factors can be solved for analytically.
Additionally, the \ac{cpd} factors of the fourth-order identity tensor in three dimensions are given by the six vertices of the regular icosahedron that lie on one side of any given plane, as shown in Fig.~\ref{fig:geometry}.
A uniform weighting factor is also associated with this geometry allowing for easy analytical solution of the weight once again.
The ansatz for these geometric solutions to the symmetric \ac{cpd} of the low-dimensional identity tensors come from the idea that the identity tensor represents spherical symmetry, and the CP factors should thus be distributed in a highly symmetric fashion.
In general, however, the weighting factors in higher dimensions are not always uniform from observation of the results of numerical optimization.

\section{Numerical Experiments}%
\label{sec:numerical_experiments}
The proposed square root second-order moment mapping is analyzed in two distinct nonlinear transformations.
The first problem consists of the mapping from Cartesian to polar coordinates and the second problem involves moment mapping in the Van der Pol oscillator system.

\subsection{Polar Transformation}%
\label{sec:polar_transformation}
The polar transformation is given by
\begin{equation}
    [r\quad \theta]^\top=\mathbf{g}([x\quad y]^\top)=\left[\sqrt{x^2+y^2}\quad \arctan\left(\frac{y}{x}\right)\right]^\top
\end{equation}
This example is representative of transformations commonly encountered in radar tracking problems during the measurement update.
The input mean and covariance are known in Cartesian coordinates with mean $\mathbf{m}_x=[0 \quad 1000]^T$ and covariance square root factor $\mathbf{S}_x=250\operatorname{diag}([4 \quad 1])$, where the $\operatorname{diag}(\cdot)$ operator returns a diagonal square matrix with its argument on the diagonal.
No noise is considered in this first experiment, which is representative of the common occurrence when one needs to simply map uncertainty between different coordinate systems.

\subsection{Van der Pol Oscillator}%
\label{sec:van_der_pol_oscillator}
In the Van der Pol problem, the discrete time system is
\begin{align}
  \mathbf{x}(t) = \mathbf{g}(\mathbf{x}_{0}) + \mathbf{w}_{0}
\end{align}
where $\mathbf{x}(t) = [x_{1} \,\, x_{2}]^\top \in \mathbb{R}^{2}$ and $\mathbf{g}(\mathbf{x}_{0})$ represents the solution flow.
The input state estimation error covariance square root factor is parameterized in terms of ratios of a reference standard deviation $\sigma$ and a rotation as
\begin{align}
  \mathbf{S}_{x} = \begin{bmatrix}
    \cos(\alpha) & -\sin(\alpha)\\
    \sin(\alpha) & \cos(\alpha)
  \end{bmatrix}
  \begin{bmatrix}
    \sigma & 0 \\
    0 & \beta \sigma
  \end{bmatrix}
\end{align}
The discrete time process noise $\mathbf{w}_{0}$ is zero mean with covariance square root factor $\mathbf{S}_{w} = \zeta \sigma \mathbf{I}^{(2)}$ where $\zeta$ is an arbitrary constant used to relate the process noise magnitude to the reference parameter $\sigma$.
The covariance parameter values for~$\sigma$,~$\beta$, and $\zeta$ are reported in Table~\ref{tab:vanderpol_parameters} and are chosen to reflect a mildly ill-conditioned covariance matrix, which can arise when one state is significantly more uncertain than another.
The solution flow is the differential equation solution
\begin{align}
  \mathbf{g}(\mathbf{x}_{0}) = \mathbf{x}_{0} + \int_{0}^{t} \dot{\mathbf{x}}(t) \mathrm{d} t
\end{align}
where
\begin{align}
  \dot{\mathbf{x}}(t) = \begin{bmatrix}
    x_{2}(t) \\
    \mu(1 - (x_{1}(t))^{2})x_{2}(t) - x_{1}(t)
  \end{bmatrix}
\end{align}
In this problem,~$\mathbf{G}^{(1)}$ and~$\mathbf{G}^{(2)}$ represent the state transition matrix and second-order state transition tensor, respectively, both of which are obtained by numerically integrating the first and second-order variational equations.
Both problems are two-dimensional, and thus the relevant \ac{cpd} factors are those of the two dimensional fourth-order identity tensor multiplied by three.
In the two-dimensional case, the factors are obtained analytically and reported in Table~\ref{tab:fourth_order_standard_gaussian_moment_cpd_factors}.
\begin{table}[htbp]
    \caption{Parameters of Van der Pol Oscillator experiment.}
    \label{tab:vanderpol_parameters}
    \centering
    \begin{tabular}{lr}
      \toprule[2pt]%
      Parameter & Value \\
      \midrule
      $\mathbf{m}_{x}$ & $[0.1 \,\, 0.5]^{\top}$\\
      $\alpha$ & $\pi/3$\\
      $\beta$ & $5\cdot10^{-6}$\\
      $\sigma$ & $10^{-1}$\\
      $\zeta$ & $10^{-3}$\\
      $\mu$ & $0.5$\\
      $(t_{0}, t_{f})$ & $(0,1)$\\
      \bottomrule
    \end{tabular}
\end{table}

\begin{table}[htbp]
    \caption{CPD factors of the bivariate standard Gaussian fourth-order moment tensor.}
    \label{tab:fourth_order_standard_gaussian_moment_cpd_factors}
    \centering
    \begin{tabular}{lrrr}
      \toprule[2pt]%
      $r$ & $\beta_{r}$ & $(v_{r})^1$ & $(v_{r})^2$\\
      \midrule
      1 & 1.63299316185545 & 1.0   & 0.0 \\
      2 & 1.63299316185545 & -0.5 & 0.866025403784439\\
      3 & 1.63299316185545 & -0.5 & -0.866025403784439 \\
      \bottomrule
    \end{tabular}
\end{table}

\subsection{Performance Analysis}%
\label{sec:performance_analysis}
The proposed second-order square root computation is, in principle, the most numerically accurate computation, with no other higher accuracy method available to serve as a ground truth or benchmark.
Instead, to assess the numerical accuracy of the square root second-order mapping, the results are compared across different floating-point precisions.
A numerically stable algorithm executed at lower precision is expected to more closely mimic the result produced using higher precision floating-point numbers than a less numerically stable algorithm.
In both examples, the second-order covariance is computed in full and square root form, as well as in single and double precision.
All comparisons are made in terms of Cholesky factors, including when comparing the covariance computed using the traditional full covariance mapping equations.
Comparing Cholesky factors instead of squaring them and comparing full covariance matrices ensures that the precision loss square root methods are designed to circumvent is not reintroduced.
In the full covariance method, all computations are performed using full covariance matrices, after which the Cholesky factor is computed directly from the final matrix $\mathbf{P}_{(2)}$.

The following notation is used to distinguish matrices computed using the different methods and precisions.
Quantities computed using single precision floating-point numbers are labeled with a ``\texttt{32}'' superscript, whereas quantities obtained using double precision are labeled with ``\texttt{64}.''
Similarly, results produced using the proposed square root method are indicated by an $\texttt{S}$ subscript and those found using the traditional full covariance equations by $\texttt{P}$.
For example, $[\mathbf{S}_{(2)}]_{\texttt{S}}^{\texttt{32}}$ is the Cholesky factor of the second-order propagated covariance using the square root formulation and computed using single precision floating-point numbers throughout the algorithm.

The differences between the full and square root computations are measured by the Frobenius norm of their square root matrix difference.
Accuracy results are reported in Tables~\ref{tab:polar_transformation_covariance_results} and~\ref{tab:vanderpol_covariance_results} for the polar transformation and Van der Pol oscillator cases, respectively.
In the polar transformation case, when using double precision floating-point numbers, the square root and full covariance results are equal to machine precision.
When reduced to single precision, the square root method matches the double precision results to within $2.5\cdot 10^{-8}$, whereas the single precision full covariance method error is three orders of magnitude greater at $6.1\cdot 10^{-5}$.

\begin{table}[htbp]
    \caption{Polar transformation covariance results.}
    \label{tab:polar_transformation_covariance_results}
    \centering
    \begin{tabular}{lr}
      \toprule[2pt]%
      Difference & Value\\
      \midrule
      $\left\Vert[\mathbf{S}_{(2)}]_{\texttt{S}}^{\texttt{64}} -
      [\mathbf{S}_{(2)}]_{\texttt{P}}^{\texttt{64}}\right\Vert_{F}$ & 0.0\\
      $\left\Vert[\mathbf{S}_{(2)}]_{\texttt{S}}^{\texttt{32}} -
      [\mathbf{S}_{(2)}]_{\texttt{P}}^{\texttt{32}}\right\Vert_{F}$ & $6.103527266532183\cdot 10^{-5}$ \\
      $\left\Vert[\mathbf{S}_{(2)}]_{\texttt{S}}^{\texttt{64}} -
      [\mathbf{S}_{(2)}]_{\texttt{S}}^{\texttt{32}}\right\Vert_{F}$ & $2.491180040031793 \cdot 10^{-8}$\\
      $\left\Vert[\mathbf{S}_{(2)}]_{\texttt{S}}^{\texttt{64}} -
      [\mathbf{S}_{(2)}]_{\texttt{P}}^{\texttt{32}}\right\Vert_{F}$ & $6.103522909335537\cdot 10^{-5}$ \\
      \bottomrule
    \end{tabular}
\end{table}

\begin{table}[htbp]
    \caption{Van der Pol oscillator covariance results.}
    \label{tab:vanderpol_covariance_results}
    \centering
    \begin{tabular}{lr}
      \toprule[2pt]%
      Difference & Value\\
      \midrule
      $\left\Vert[\mathbf{S}_{(2)}]_{\texttt{S}}^{\texttt{64}} -
      [\mathbf{S}_{(2)}]_{\texttt{P}}^{\texttt{64}}\right\Vert_{F}$ &  $ 7.05981306231723\cdot 10^{-17}$\\
      $\left\Vert[\mathbf{S}_{(2)}]_{\texttt{S}}^{\texttt{32}} -
      [\mathbf{S}_{(2)}]_{\texttt{P}}^{\texttt{32}}\right\Vert_{F}$ & $2.2371136054744056 \cdot 10^{-8}$ \\
      $\left\Vert[\mathbf{S}_{(2)}]_{\texttt{S}}^{\texttt{64}} -
      [\mathbf{S}_{(2)}]_{\texttt{S}}^{\texttt{32}}\right\Vert_{F}$ & $5.246760444776802 \cdot 10^{-10}$\\
      $\left\Vert[\mathbf{S}_{(2)}]_{\texttt{S}}^{\texttt{64}} -
      [\mathbf{S}_{(2)}]_{\texttt{P}}^{\texttt{32}}\right\Vert_{F}$ & $2.2380985921863992\cdot 10^{-8}$ \\
      \bottomrule
    \end{tabular}
\end{table}
Similar trends are observed in the Van der Pol oscillator experiment, where when using double precision floating point numbers, the full and square root computations match to within $7.1\cdot 10^{-17}$.
The square root method is again shown to be numerically stable, matching the double precision result to within $5.3\cdot10^{-10}$ when using single precision floating numbers.
On the other hand, the single precision full covariance computation is two orders of magnitude higher in error compared to square root computation.

When using single precision floating-point arithmetic, the advantages of the square root computation over its full covariance counterpart are clear.
While at first consideration, using double precision floating-point arithmetic might remediate the worst numerical errors, even small errors can accumulate over time while remaining undetected, degrading filter performance or in some cases leading to catastrophic failure \cite{thornton1976NumericalComparisonDiscrete}.
Furthermore, the square root method guarantees symmetry and positive definiteness by construction whereas full covariance computations are known to violate these conditions numerically.

\acresetall
\section{Conclusion}
\label{sec:conclusion}
This work developed a square root form of the second-order approximation of mean and covariance mapping of a multivariate Gaussian distribution through a nonlinear function.
The fundamental building blocks enabling this development are standard square root state estimation techniques involving the Cholesky and QR decomposition along with a novel methodology based on whitening transformations and the \ac{cpd} of the standard multivariate Gaussian fourth-order central moment tensor.
The methodology presented here has two major benefits over standard second-order moment mapping.
First, this method inherently ensures that issues with symmetric positive definiteness are never encountered in uncertainty mapping.
Similarly, numerical precision of the covariance is also improved.
The square root approach demonstrated orders of magnitude more accurate solutions compared to the traditional approach in two distinct nonlinear examples involving the polar transformation and Van der Pol oscillator.
Second, this methodology is computationally more efficient than the original second-order covariance mapping method, requiring fewer floating point operations.
The only limitation of this method is that it requires the original distribution to be Gaussian, though for uncertainty propagation and state estimation methods this is often a tacit assumption in state-of-the-art methods when higher-order moments of a distribution are not explicitly tracked.

\section*{Acknowledgments}
This material is based upon work supported by the Air Force Office of Scientific Research under award number FA9550-25-1-0101.
\bibliography{references}

\end{document}